\documentclass[sigconf, screen]{acmart}
\AtBeginDocument{%
  }
\setcopyright{acmlicensed}
\copyrightyear{2025}
\acmYear{2025}
\acmDOI{XXXXXXX.XXXXXXX}
\acmConference[MM '25]{ACM International Conference on Multimedia 2025}{October 27--31,
  2025}{Dublin, Ireland}
\acmISBN{978-1-4503-XXXX-X/2018/06}

\definecolor{customcite}{HTML}{e67a7a}
\definecolor{customlink}{HTML}{b83b5e}
\definecolor{customurl}{HTML}{11999e}
\AtEndPreamble{
\usepackage{hyperref}

    \hypersetup{
      colorlinks = true,
      linkcolor = customlink,
      anchorcolor = purple,
      citecolor = customcite,
      filecolor = purple,
      urlcolor = customurl
    }
}

\usepackage{tabularx}

\begin{document}

\title{HyperFusion: Hierarchical Multimodal Ensemble Learning for Social Media Popularity Prediction}
\author{Liliang Ye}
\affiliation{%
  \institution{Huazhong University of Science and Technology}
  \city{Wuhan}
  \country{China}
}
\email{yll@hust.edu.cn}

\author{Yunyao Zhang}
\affiliation{%
  \institution{Huazhong University of Science and Technology}
  \city{Wuhan}
  \country{China}
}
\email{ikoyun@hust.edu.cn}

\author{Yafeng Wu}
\affiliation{%
 \institution{Huazhong University of Science and Technology}
 \city{Wuhan}
 \country{China}
 }
\email{wyf2024@hust.edu.cn}

\author{Yi-Ping Phoebe Chen}
\affiliation{%
 \institution{La Trobe University}
 \city{Melbourne}
 \country{Australia}
 }
\email{phoebe.chen@latrobe.edu.au}

\author{Junqing Yu}
\affiliation{%
 \institution{Huazhong University of Science and Technology}
 \city{Wuhan}
 \country{China}
 }
\email{yjqing@hust.edu.cn}

\author{Wei Yang}
\affiliation{%
 \institution{Huazhong University of Science and Technology}
 \city{Wuhan}
 \country{China}
 }
\email{weiyangcs@hust.edu.cn}

\author{Zikai Song}
\authornote{Corresponding author.}
\affiliation{%
  \institution{Huazhong University of Science and Technology}
  \city{Wuhan}
  \country{China}
}
\email{skyesong@hust.edu.cn}

\renewcommand{\shortauthors}{Liliang Ye et al.}

\begin{abstract}
  Social media popularity prediction plays a crucial role in content optimization, marketing strategies, and user engagement enhancement across digital platforms. However, predicting post popularity remains challenging due to the complex interplay between visual, textual, temporal, and user behavioral factors. This paper presents HyperFusion, a hierarchical multimodal ensemble learning framework for social media popularity prediction. Our approach employs a three-tier fusion architecture that progressively integrates features across abstraction levels: visual representations from CLIP encoders, textual embeddings from transformer models, and temporal-spatial metadata with user characteristics. The framework implements a hierarchical ensemble strategy combining CatBoost, TabNet, and custom multi-layer perceptrons. To address limited labeled data, we propose a two-stage training methodology with pseudo-labeling and iterative refinement. We introduce novel cross-modal similarity measures and hierarchical clustering features that capture inter-modal dependencies. Experimental results demonstrate that HyperFusion achieves competitive performance on the SMP challenge dataset. Our team achieved third place in the SMP Challenge 2025 (Image Track). The source code is available at \url{https://anonymous.4open.science/r/SMPDImage}.
\end{abstract}



\maketitle

\section{INTRODUCTION}
\label{introduction}

The exponential growth of social media platforms has fundamentally transformed digital communication, with billions of posts shared daily on diverse platforms around the world~\cite{GlobalSocialMedia}. Predicting the popularity of social media content has emerged as a critical research area with profound implications for content optimization, marketing strategies, and algorithmic recommendation systems~\cite{jeong2024enhancing}. The complexity of this task stems from the intricate interplay between multiple heterogeneous factors, including visual aesthetics, textual semantics, temporal dynamics, spatial context, and user behavioral patterns. Early research~\cite{wuSequentialPredictionSocial2017a, wu2016unfolding, meghawat2018multimodal, wu2023smp, laiHyFeaWinningSolution2020, wangFeatureGeneralizationFramework2020} primarily relied on handcrafted features derived from textual content and basic metadata such as posting timestamps and user demographics. While these approaches provided foundational insights, they often failed to capture the rich semantic information embedded in visual content and the complex cross-modal relationships that significantly influence user engagement. The advent of deep learning has revolutionized this field, enabling researchers to leverage sophisticated neural architectures for extracting meaningful representations from multimodal data~\cite{chenDoubleFineTuningMultiObjectiveVisionandLanguage2023a, chengRetrievalAugmentedHypergraphMultimodal2024, zhangContrastiveLearningImplicit2024, linMMFWinningSolution2024}. Recent advances have demonstrated the effectiveness of vision-language models, particularly CLIP~\cite{radford2021learning}, in bridging the semantic gap between visual and textual modalities. However, existing approaches often treat different modalities independently or employ simple concatenation strategies for feature fusion, potentially overlooking the hierarchical nature of feature interactions and the varying modality importance across contexts.

To address these limitations, this paper introduces HyperFusion, a novel hierarchical multimodal ensemble learning framework specifically designed for social media popularity prediction. Our approach fundamentally differs from existing methods by implementing a three-tier hierarchical fusion architecture that progressively integrates features across multiple abstraction levels. The framework combines visual representations extracted from CLIP encoders, textual embeddings from transformer-based models~\cite{vaswani2017attention}, and comprehensive temporal-spatial metadata with user characteristics through a sophisticated hierarchical ensemble strategy. The key innovation of HyperFusion lies in its ability to automatically learn the optimal fusion weights at different hierarchical levels while simultaneously addressing the data scarcity problem through a two-stage training methodology incorporating pseudo-labeling and iterative refinement. Additionally, we introduce novel cross-modal similarity measures and hierarchical clustering features that effectively capture inter-modal dependencies, enabling more nuanced understanding of content popularity patterns.

Our comprehensive experimental evaluation on the SMP Challenge dataset~\cite{wu2024smp} demonstrates that HyperFusion achieves competitive performance, ranking \textbf{third place} in the Image Track. This result provides valuable insights into the relative importance of different modalities and feature types in determining social media content popularity. The hierarchical fusion approach consistently outperforms traditional concatenation methods, validating the effectiveness of our proposed architecture in handling complex multimodal interactions.

\section{RELATED WORK}
\label{related-work}

Social media popularity prediction has evolved from early handcrafted feature approaches to sophisticated multimodal learning\cite{song2024autogenic,luo2024diffusiontrack,li2024coupled,hu2025sf2t} frameworks. Initial studies~\cite{zhang2018user, meghawat2018multimodal} focused on extracting features from textual descriptions, temporal patterns, and user demographics, establishing the fundamental understanding that engagement prediction requires comprehensive modeling of heterogeneous information sources. The transition to deep learning methodologies has fundamentally transformed this landscape, with contemporary approaches~\cite{huDualStreamPreTrainingTransformer2024, hsuRevisitingVisionLanguageFeatures2024, mao2023enhanced} leveraging powerful vision models~\cite{song2022transformer,song2023compact,zhou2025video,song2025temporal} or vision-language models such as CLIP and BERT~\cite{devlin2019bert} to extract rich semantic representations from both visual and textual content. These pre-trained encoders enable unified understanding of multimodal posts, capturing subtle engagement cues that traditional feature engineering methods often overlook.

The integration of multimodal information has become increasingly sophisticated through advanced fusion architectures. Hierarchical and attention-based mechanisms have shown particular effectiveness in modeling complex cross-modal interactions~\cite{wang2023social}, while ensemble learning strategies have emerged as essential components of state-of-the-art solutions. Modern frameworks typically combine gradient boosting methods including XGBoost~\cite{chen2016xgboost}, CatBoost~\cite{dorogush2018catboost}, and LightGBM~\cite{ke2017lightgbm} with neural network architectures such as TabNet~\cite{arik2021tabnet} and multilayer perceptrons, capitalizing on the complementary strengths of tree-based models in handling heterogeneous features and neural networks in capturing non-linear relationships. In addition to architectural improvements, data-centric strategies have also gained attention. To address data scarcity challenges inherent in social media datasets, recent works~\cite{kang2019catboost, lin2021social, lin2024mmf} have increasingly adopted pseudo-labeling strategies and iterative training procedures that effectively utilize unlabeled data while improving model robustness.

Despite these advances, several fundamental challenges persist in social media popularity prediction. The highly skewed distribution of engagement scores, noisy or incomplete metadata, and the dynamic nature of user preferences continue to complicate model development and evaluation. These challenges have driven the development of more robust feature extraction pipelines, sophisticated ensemble methodologies, and adaptive training strategies that can handle the inherent uncertainty and variability in social media data. The convergence of these methodological advances has established a clear paradigm emphasizing multimodal feature integration, powerful pre-trained representation models, and ensemble learning strategies as the foundation for effective social media popularity prediction systems.
\section{METHODOLOGY}
\label{methodology}

\begin{figure*}[h!]
    \centering
    \includegraphics[width=\linewidth]{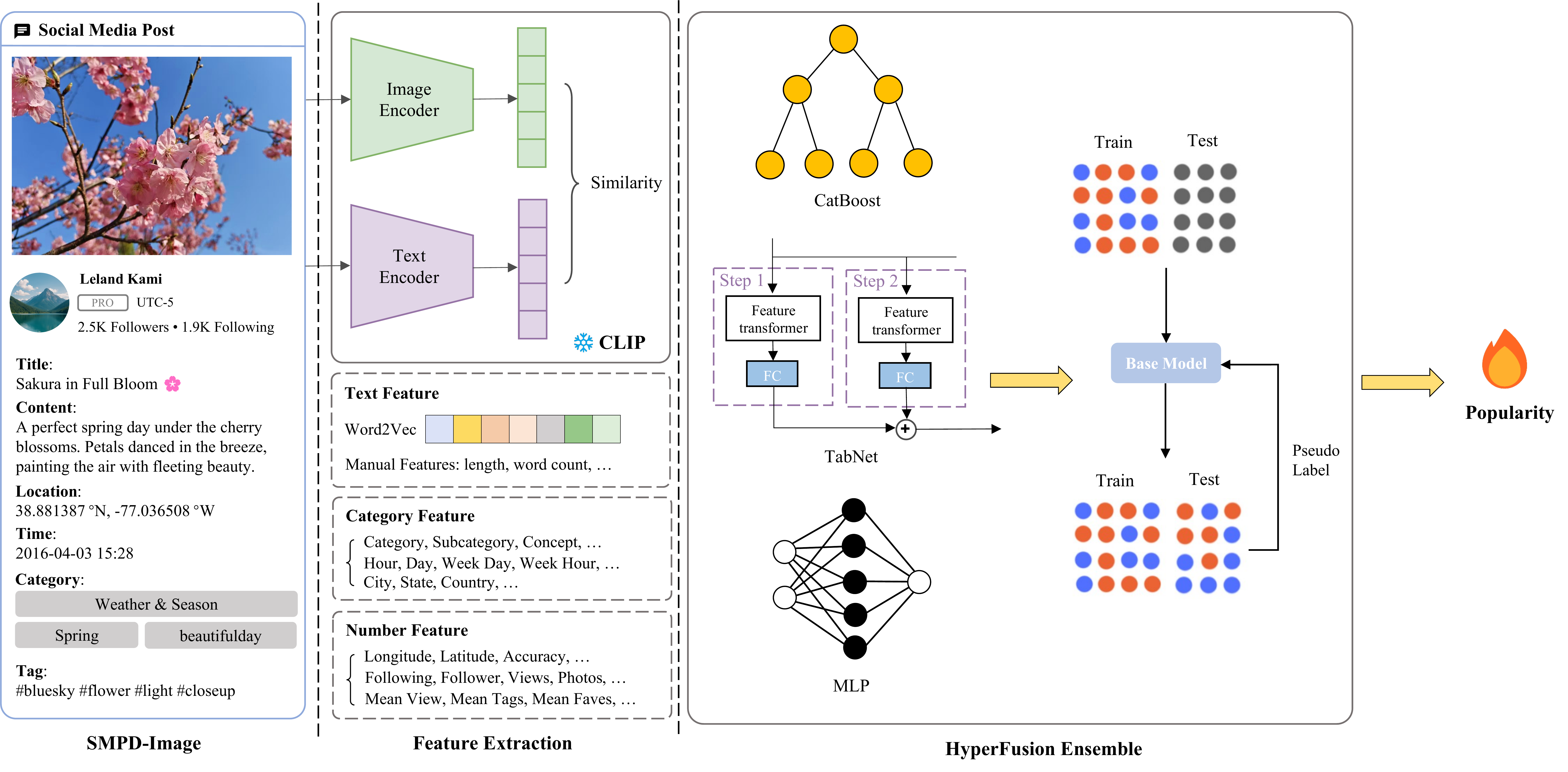} 
    \caption{The HyperFusion framework pipeline integrates visual, textual, spatiotemporal, and user features through hierarchical ensemble learning for multimodal prediction of social media content popularity.}
    \label{fig:framework}
\end{figure*}

\subsection{Overview}
In this section, we present the design of the HyperFusion framework, which systematically addresses the challenge of predicting social media popularity through comprehensive multimodal analysis. The architecture of HyperFusion is specifically tailored to integrate heterogeneous sources of information and model complex interactions across multiple modalities. The pipeline begins with the extraction of high-level semantic representations from visual content using pretrained encoders, followed by the integration of textual features, user characteristics, and spatiotemporal signals. To ensure feature compatibility and minimize the influence of noise, we implement a series of preprocessing steps, including normalization, outlier removal, and dimensionality reduction. The resulting multimodal feature set is subsequently processed through an ensemble of specialized models, each capable of capturing distinct aspects of the prediction task. The overall framework is designed to be robust and interpretable, facilitating reliable modeling of the diverse and dynamic nature of social media content. In the following subsections, we describe each component of the pipeline in detail.

\subsection{Problem Formulation}
We formulate social media popularity prediction as a multimodal regression problem that leverages comprehensive feature representations. Given a social media post $p$ with associated metadata, our objective is to predict its popularity score $y \in \mathbb{R}^+$ through the mapping function:
\begin{equation}
y = f(\mathbf{v}, \mathbf{t}, \mathbf{s}, \mathbf{u}, \mathbf{r}; \theta)
\end{equation}
where $\mathbf{v} \in \mathbb{R}^{d_v}$, $\mathbf{t} \in \mathbb{R}^{d_t}$, $\mathbf{s} \in \mathbb{R}^{d_s}$, $\mathbf{u} \in \mathbb{R}^{d_u}$, and $\mathbf{r} \in \mathbb{R}^{d_r}$ represent visual, textual, spatiotemporal, user, and cross-modal similarity feature vectors respectively, with $\theta$ denoting the learnable parameters of the prediction model.

\subsection{Multimodal Feature Construction and Integration}
A central aspect of the HyperFusion framework is the systematic extraction and integration of multimodal features that capture the diverse factors influencing social media popularity. We organize feature engineering into five principal components: visual features, textual features, spatiotemporal features, user characteristics, and cross-modal coherence measures, each offering distinct perspectives on content engagement potential.

\textbf{Visual Feature Extraction.}
We employ a pretrained CLIP encoder to extract deep visual representations from each image. The CLIP visual encoder processes input images to generate high dimensional embeddings that capture semantic visual content. Specifically, let $\mathbf{I} \in \mathbb{R}^{H \times W \times 3}$ denote an input image, where $H$ and $W$ represent height and width respectively. The visual feature $\mathbf{v}_{visual}$ is computed as:
\begin{equation}
\mathbf{v}_{visual} = \text{CLIP}_{visual}(\mathbf{I})
\end{equation}
where $\text{CLIP}_{visual}(\cdot)$ represents the CLIP visual encoder that outputs a 512-dimensional feature vector. To further enhance representational capacity while maintaining computational efficiency, we apply Principal Component Analysis (PCA)~\cite{hotelling1933analysis} for dimensionality optimization when necessary.

\textbf{Textual Feature Extraction.}
Our textual feature extraction combines multiple approaches to capture both semantic and statistical properties of text content. We utilize CLIP text encoders for cross-modal semantic alignment and GloVe~\cite{pennington2014glove} embeddings for traditional word-level representations. For hashtags and titles, we generate 300-dimensional feature vectors using GloVe embeddings. Let $\mathbf{T} = \{w_1, w_2, \ldots, w_n\}$ represent a text sequence with $n$ words. The textual feature $\mathbf{v}_{textual}$ is constructed as:
\begin{equation}
\mathbf{v}_{textual} = [\text{CLIP}_{text}(\mathbf{T}); \text{GloVe}(\mathbf{T}); \mathbf{f}_{stat}(\mathbf{T})]
\end{equation}
where $\text{CLIP}_{text}(\cdot)$ generates semantic embeddings, $\text{GloVe}(\cdot)$ produces word-level representations, and $\mathbf{f}_{stat}(\cdot)$ extracts statistical features including text length, word count, and linguistic patterns.

\textbf{Spatiotemporal and User Modeling.}
Temporal dynamics are captured through posting timestamps, account age calculations, and seasonal patterns that reflect content virality trends. Geographic context is encoded via coordinate information and location accuracy measures. User characteristics are modeled through behavioral patterns, engagement history, and professional status indicators. We apply Singular Value Decomposition (SVD) to user interaction matrices to generate compact user and location embeddings. Formally, given a user-item interaction matrix $\mathbf{M} \in \mathbb{R}^{m \times n}$, the SVD decomposition yields:
\begin{equation}
\mathbf{M} = \mathbf{U}\mathbf{\Sigma}\mathbf{V}^T
\end{equation}
where $\mathbf{U} \in \mathbb{R}^{m \times k}$ provides 399-dimensional user embeddings and location embeddings of 400 dimensions are derived from $\mathbf{V} \in \mathbb{R}^{n \times k}$, capturing latent behavioral and geographic patterns.

\textbf{Cross-Modal Coherence Measurement.}
To quantify the semantic alignment between visual and textual content, we compute CLIP-based similarity scores that measure multimodal consistency. Given visual features $\mathbf{v}_{visual}$ and textual features $\mathbf{v}_{textual}$, the cross-modal similarity is computed as:
\begin{equation}
s_{cross} = \frac{\mathbf{v}_{visual} \cdot \mathbf{v}_{textual}}{\|\mathbf{v}_{visual}\| \|\mathbf{v}_{textual}\|}
\end{equation}
These coherence measures provide crucial insights into content consistency, which significantly influences user engagement and sharing behavior.

\textbf{Feature Integration and Preprocessing.}
We concatenate all extracted features to form a unified multimodal feature vector for each post. Formally, let $\mathbf{v}_{visual}$, $\mathbf{v}_{textual}$, $\mathbf{v}_{spatial}$, $\mathbf{v}_{user}$, and $\mathbf{v}_{cross}$ represent the feature vectors from different modalities. The final multimodal feature vector $\mathbf{x}$ is constructed as:
\begin{equation}
\mathbf{x} = [\mathbf{v}_{visual}; \mathbf{v}_{textual}; \mathbf{v}_{spatial}; \mathbf{v}_{user}; \mathbf{v}_{cross}]
\end{equation}
where $[\cdot; \cdot]$ denotes the concatenation operation. Before feeding the features into the models, we implement comprehensive preprocessing including missing value imputation with context-appropriate defaults, outlier filtering using the interquartile range method, and feature normalization to ensure stable distributions across different modalities.

\subsection{HyperFusion Ensemble Architecture}
Our HyperFusion framework implements a sophisticated ensemble strategy that combines multiple specialized models, each designed to capture different aspects of the multimodal feature space and prediction task.

\textbf{Model Components.}
The ensemble incorporates four complementary architectures, each optimized for specific characteristics of the multimodal data. CatBoost Regressor serves as the primary gradient boosting component, specifically designed for structured data with categorical features. It utilizes category-aware encoding strategies and handles missing values naturally. TabNet provides interpretable deep tabular learning with sequential attention mechanisms that enable automatic feature selection and provide insights into feature importance. A custom Multi-Layer Perceptron (MLP) with co-attention layers captures complex visual-textual interactions through learned attention weights, enabling fine-grained cross-modal understanding. The CLIP-based Hierarchical Predictor leverages pretrained multimodal representations through multiple fully connected layers with ReLU activation and batch normalization.

\textbf{Training Objective and Optimization.}
For the ensemble training, we employ a robust loss function that balances prediction accuracy with outlier resistance. The training objective for each model $f_i$ is formulated using the Huber loss:
\begin{equation}
\mathcal{L}_{\delta}(y, \hat{y}) = \begin{cases}
\frac{1}{2}(y - \hat{y})^2, & \text{if } |y - \hat{y}| \leq \delta \\
\delta |y - \hat{y}| - \frac{1}{2}\delta^2, & \text{otherwise}
\end{cases}
\end{equation}
where $y$ denotes the ground-truth popularity score, $\hat{y}$ represents the predicted value, and $\delta$ determines the transition point between quadratic and linear behavior, thereby improving robustness to outliers while maintaining sensitivity to regular samples.

\textbf{Ensemble Integration Strategy.}
The final prediction combines individual model outputs through optimized weighted averaging. We apply five-fold cross-validation by splitting the training set into five subsets and iteratively using four for training and one for validation. For each fold $k$, each model $f_i^{(k)}$ is trained independently. The ensemble prediction for fold $k$ is computed as:
\begin{equation}
\hat{y}^{(k)} = \sum_{i=1}^{N} w_i^{(k)} \cdot f_i^{(k)}(\mathbf{x})
\end{equation}
The final prediction $\hat{y}$ is calculated by averaging the outputs across all folds:
\begin{equation}
\hat{y} = \frac{1}{K} \sum_{k=1}^{K} \hat{y}^{(k)}
\end{equation}
where $K=5$ represents the number of folds, and weights $w_i^{(k)}$ are optimized through cross-validation to maximize ensemble performance on validation data.

\subsection{Semi-Supervised Learning Strategy}
We employ a sophisticated two-stage training approach that maximizes both model performance and generalization capability through iterative pseudo-labeling and semi-supervised learning techniques.

\textbf{Base Model Training.}
In the initial stage, individual models are trained on the original labeled dataset using stratified cross-validation for robust hyperparameter optimization. Each model specializes in different aspects of the multimodal feature space: CatBoost focuses on categorical relationships and non-linear patterns inherent in user behavior and metadata, TabNet provides interpretable attention-based learning with automatic feature selection capabilities, and the MLP captures complex multimodal interactions through sophisticated co-attention mechanisms that model visual-textual dependencies.

\textbf{Iterative Pseudo-Label Enhancement.}
The second stage implements an iterative pseudo-labeling strategy to effectively leverage unlabeled test data. High-confidence ensemble predictions are systematically incorporated as pseudo-labels for additional training iterations. This semi-supervised approach updates the training dataset by selecting predictions with confidence scores above empirically determined thresholds $\tau$, defined as:
\begin{equation}
\tau = \mu_{confidence} + \alpha \cdot \sigma_{confidence}
\end{equation}
where $\mu_{confidence}$ and $\sigma_{confidence}$ represent the mean and standard deviation of prediction confidence scores, and $\alpha$ is a hyperparameter controlling the selection strictness. This strategy enables the model to learn from broader data distributions while maintaining prediction quality and avoiding the incorporation of low-quality pseudo-labels that could degrade performance.

\section{EXPERIMENT}
\label{experiment}

\subsection{Dataset}
The SMP dataset comprises 486,000 image posts from 70,000 users over a 16-month period, encompassing a wide range of content categories~\cite{wu2019smp}. Each post is associated with high-resolution images, textual descriptions, posting time, geographic information, and user profiles. The dataset supports multimodal popularity prediction by providing aligned visual, textual, spatiotemporal, and user-related features. Popularity scores are continuous values derived from real user engagement, enabling regression-based modeling and evaluation.

\subsection{Evaluation Metrics}
To comprehensively evaluate model performance, we adopt two metrics: Spearman’s Rank Correlation (SRC) and Mean Absolute Error (MAE). SRC assesses the monotonic relationship between predicted and actual popularity, reflecting the model’s ability to preserve correct ranking among samples. The SRC is defined as:
\begin{equation}
\mathrm{SRC} = 1 - \frac{6 \sum_{i=1}^{n} d_i^2}{n(n^2-1)}
\end{equation}
where $d_i$ is the difference between the ranks of the $i$-th predicted and ground-truth values. MAE measures the average absolute difference between predicted and ground-truth values, providing an interpretable error in the original scale:
\begin{equation}
\mathrm{MAE} = \frac{1}{n} \sum_{i=1}^{n} |y_i - \hat{y}_i|
\end{equation}
This dual-metric evaluation ensures both ranking fidelity and numerical accuracy are considered in the assessment of popularity prediction models.

\subsection{Main Results}

\subsubsection{Overall Performance}
Our model achieved strong results on the official evaluation dataset, with a Spearman's rank correlation coefficient (SRC) of 0.7324 and a mean absolute error (MAE) of 1.2402. These results reflect the effectiveness of our multimodal feature fusion and model design. The approach generalizes well across diverse samples and demonstrates robust performance in popularity prediction tasks.

\subsubsection{Ablation Study}

To further elucidate the contribution of each modality and the ensemble strategy, we conduct a focused ablation study. ~\autoref{tab:ablation} summarizes the results for the following configurations: the full model, and variants with each major modality or the ensemble mechanism removed. The evaluation is performed on the official validation set, and both SRC and MAE are reported for each setting.

\begin{table}[!htbp]
    \centering
    \caption{Ablation study results on the validation set.}
    \begin{tabularx}{0.85\linewidth}{X c c}
        \toprule[1.2pt]
        \textbf{Methods} & \textbf{SRC} & \textbf{MAE} \\
        \midrule
        Full Model & 0.7324 & 1.2402 \\
        w/o Visual & 0.6916 & 1.3503 \\
        w/o Metadata & 0.7068 & 1.3441 \\
        w/o Textual & 0.7196 & 1.2663 \\
        w/o Geolocation & 0.7294 & 1.2445 \\
        w/o Ensemble & 0.7253 & 1.2631 \\
        w/o Pseudo Label & 0.7287 & 1.2774 \\
        w/o Filter & 0.7249 & 1.2785 \\
        w/o 5-Fold & 0.7290 & 1.2773 \\
        \bottomrule[1.2pt]
    \end{tabularx}
    \label{tab:ablation}
\end{table}

Our full model achieves the best performance, confirming the benefits of integrating multiple modalities and advanced training strategies. We observe a significant performance drop when removing visual features, which underscores the importance of visual content in predicting popularity. Similarly, excluding textual and metadata features also degrades performance, highlighting their complementary roles in providing semantic and contextual information. The removal of geolocation data results in a minor accuracy decrease, suggesting its auxiliary yet valuable contribution.

We also investigate the impact of our proposed training techniques. Removing the pseudo-labeling stage leads to a noticeable performance decline, which validates the effectiveness of our semi-supervised approach in leveraging unlabeled data. The exclusion of the outlier filtering process also impairs results, confirming that robust data preprocessing is critical for model stability. Furthermore, forgoing the 5-fold cross-validation strategy for a single training split results in a substantial performance drop, which highlights the importance of cross-validation for mitigating overfitting and enhancing generalization. These results collectively affirm that each component of our framework, from feature engineering to training methodology, is integral to achieving strong predictive performance.

\subsubsection{Feature Importance}
To further elucidate the predictive mechanism of our model, we conduct a comprehensive feature importance analysis based on the averaged results across multiple runs. ~\autoref{tab:importance} presents the top 20 most influential features. User-centric attributes, such as user identity, posting activity, and follower statistics, consistently dominate the ranking, highlighting the pivotal role of user information in popularity prediction. Semantic features, including concept category and tag-related statistics, also contribute substantially, reflecting the importance of content semantics. Visual and temporal cues, as well as latent representations from user and tag embeddings, provide complementary perspectives that enhance model robustness. The diversity among the top features underscores the necessity of integrating heterogeneous modalities within a unified framework.

\begin{table}[!htbp]
    \centering
    \caption{Importance of different features.}
    \resizebox{1\linewidth}{!}{%
    \begin{tabular}{llc|llc}
        \toprule[1.2pt]
        \textbf{Rank} & \textbf{Feature} & \textbf{Importance} & \textbf{Rank} & \textbf{Feature} & \textbf{Importance} \\
        \midrule
        1  & User ID & 8.85 & 11 & Total Tags & 0.67 \\
        2  & Concept & 6.80 & 12 & Group Count & 0.63 \\
        3  & Post Count & 4.81 & 13 & Mean Favorites & 0.56 \\
        4  & Follower Count & 4.33 & 14 & Tag Vec 218 & 0.52 \\
        5  & Total Views & 3.28 & 15 & Tag Vec 15 & 0.51 \\
        6  & Photo Count & 2.66 & 16 & User Vec 18 & 0.51 \\
        7  & Tag Number & 2.25 & 17 & First Week Taken & 0.48 \\
        8  & First Date Posted & 0.89 & 18 & User Vec 16 & 0.43 \\
        9  & Mean Tag & 0.84 & 19 & Following Count & 0.42 \\
        10 & First Date Taken & 0.80 & 20 & Tag Vec 224 & 0.42 \\
        \bottomrule[1.2pt]
    \end{tabular}}
    \label{tab:importance}
\end{table}

These findings confirm that user-centric and semantic features are indispensable for accurate popularity estimation, while visual and temporal signals provide valuable complementary information. The integration of diverse modalities enables the model to capture the multifaceted nature of social media dynamics, leading to robust and generalizable predictions.

\subsection{Analysis}
\begin{figure}[h!]
    \centering
    \includegraphics[width=1\linewidth]{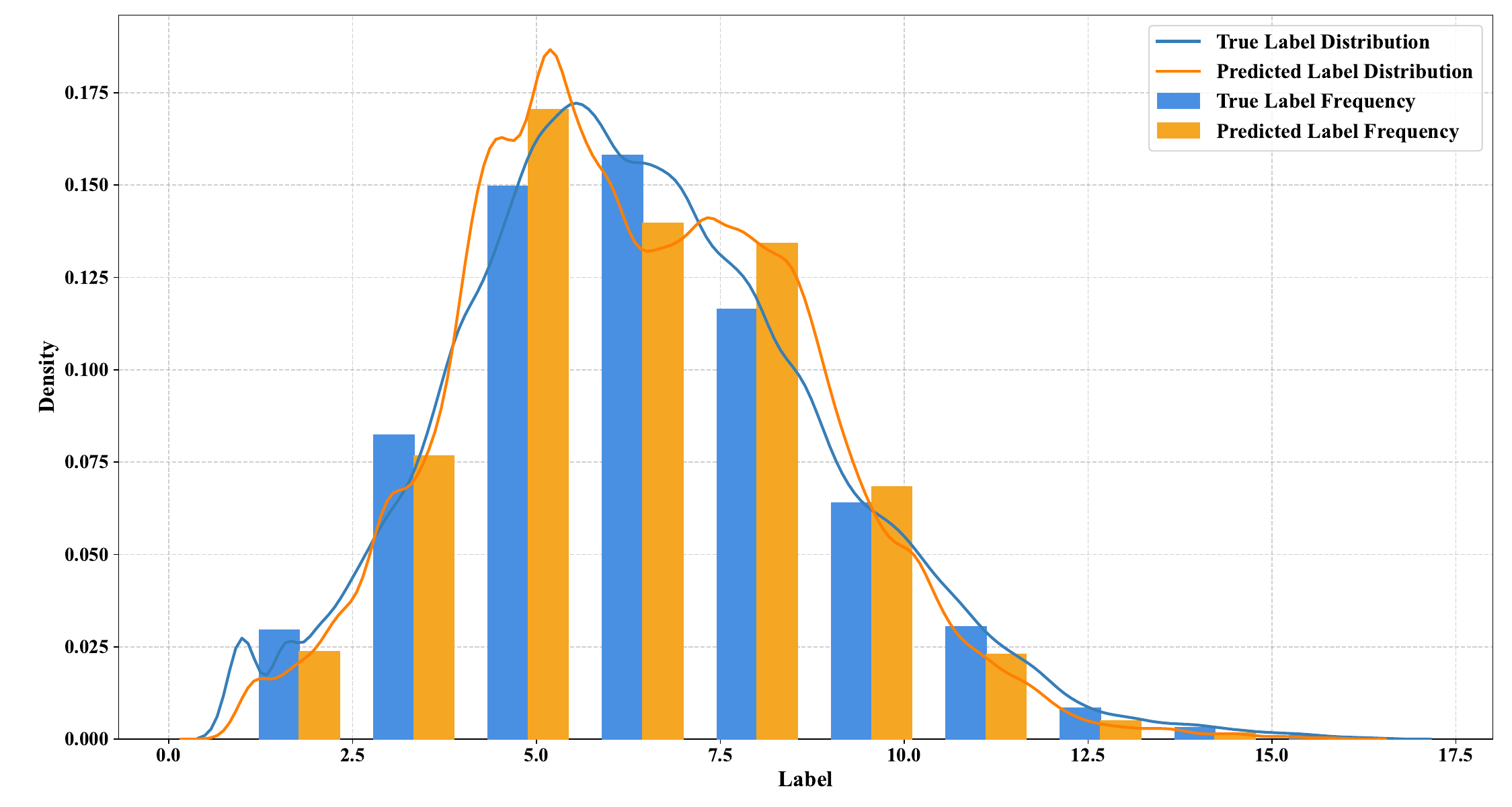}
    \caption{Histogram and kernel density estimation of predicted vs. ground-truth label distributions on the validation set.}
    \label{fig:label_density}
\end{figure}

To gain deeper insight into the predictive behavior of our model, we analyze the distribution of predicted and true popularity scores on the validation set. As illustrated in ~\autoref{fig:label_density}, both the histogram and density curves reveal that the predicted scores closely follow the empirical distribution of the ground-truth labels. The model successfully captures the unimodal and right-skewed nature of the data, with the highest density concentrated between 5 and 10. This alignment demonstrates that the model is able to learn the dominant statistical patterns present in real-world popularity data.

A moderate smoothing effect is observed in the predicted distribution, which can be attributed to the ensemble inference and regularization strategies that promote generalization. While the model achieves reliable calibration across the majority of the label range, there is a tendency for predictions to be more conservative at the lower and upper extremes. In particular, the frequency of predicted scores in the lowest and highest intervals is slightly reduced compared to the true distribution, reflecting the challenge posed by data imbalance in these regions.

Despite these limitations, the overall consistency between predicted and actual distributions underscores the robustness of the proposed approach for large-scale popularity estimation. The model produces scores with meaningful variance and stable calibration, supporting its practical applicability in realistic social media scenarios. Future work may explore advanced techniques such as label distribution adjustment or targeted loss re-weighting to further improve performance on underrepresented cases.

\section{CONCLUSION AND FUTURE WORK}
\label{conclusion-and-future-work}

This work presents HyperFusion, a comprehensive hierarchical multimodal ensemble learning framework for social media popularity prediction that systematically integrates visual, textual, spatiotemporal, and user behavioral information. Through careful orchestration of CLIP-based visual encoders, transformer-derived textual embeddings, and sophisticated ensemble strategies, our approach effectively captures the complex multimodal patterns underlying social engagement dynamics. The experimental evaluation demonstrates that the proposed hierarchical fusion architecture achieves competitive performance with an SRC of 0.7324 and MAE of 1.2402, validating the effectiveness of progressive feature integration across multiple abstraction levels.

The comprehensive ablation studies reveal that each modality contributes meaningfully to the overall predictive performance, with visual features providing essential aesthetic and semantic cues, textual components capturing linguistic patterns and content semantics, and user characteristics offering crucial behavioral insights. The feature importance analysis further confirms that user-centric attributes dominate the predictive landscape, while semantic features and cross-modal coherence measures provide valuable complementary information. These findings highlight the importance of comprehensive multimodal modeling for accurate popularity prediction.

Nonetheless, several challenges remain for future exploration. The dynamic nature of social media trends, potential semantic inconsistencies across modalities, and the computational complexity of hierarchical fusion present ongoing research opportunities. Future investigations may focus on developing more adaptive fusion mechanisms that can dynamically adjust to evolving content patterns, as well as exploring temporal modeling approaches that capture the evolution of popularity over time. Moreover, enhancing the interpretability of cross-modal interactions and developing more efficient training strategies for large-scale deployment represent promising research directions.

In summary, this study contributes to the understanding of hierarchical multimodal learning in social media analysis and offers insights that may inform the development of more effective content popularity prediction systems across digital platforms.


\bibliographystyle{ACM-Reference-Format}
\bibliography{main}

\end{document}